\documentstyle[referee]{laa}     

\def\h{$h^{-1}$ Mpc}

\begin{document}

\thesaurus{12(11.03.1; 12.12.1)}

\title{The spatial distribution of nearby galaxy clusters
in the northern and southern galactic hemispheres}

\author{A.~Cappi\inst{1,2} \and S.~Maurogordato\inst{2}}

\offprints{S.~Maurogordato}

\institute{Osservatorio astronomico di Bologna, via Zamboni 33, I-40126
Bologna, Italy
\and
Centre National de la Recherche Scientifique,
DAEC Observatoire de Meudon, Place J.Janssen, 92195
 Meudon France}

\date{Received December 23, 1991; accepted February 20, 1992}

   \maketitle

   \begin{abstract}
%
%
We compare the spatial distributions of galaxy clusters in the
northern and southern galactic hemispheres, and
the Abell and ACO clusters distributions.
We perform a statistical (correlation and cluster) analysis
of a sample of Abell and ACO galaxy clusters in the
southern galactic hemisphere ($b_{II} \le -40\degr$ and $m_{10} \le 16.5$).
We compare these results with a symmetric sample
($b_{II} \ge +40\degr$ and $m_{10} \le 16.5$) at
northern galactic latitude taken from Postman et al. (1992).
For the northern sample, we substantially confirm the
results of Postman et al.
We find that the two-point spatial correlation function $\xi_{cc}(r)$ of
northern and southern clusters is comparable, with mean correlation
length $r_0 \sim $
19.6 \h ~and slope $\gamma \sim$ -1.8. Moreover, ACO and Abell clusters show
similar spatial correlations.
$\xi_{cc}(r)$ is positive up to $\sim$ 45 \h ~in all our
samples, and it is systematically negative in the range $50 \la r \la 100$ \h.
Percolation properties are remarkably similar in the northern
and southern cluster samples. We give also a catalog of superclusters. In
the south galactic hemisphere the main feature is a very rich, extended
supercluster (spreading over $\sim$ 65 \h) in the Horologium region
at a redshift $z \sim 0.06$, near to a large void.

   \end{abstract}

%

\section{Introduction}

Galaxy cluster catalogs offer us the possibility to
study the large-scale structure of the Universe
in much larger volumes than presently available galaxy
catalogs, reaching depths beyond $z \sim 0.2$ (see Huchra et al. 1990).
Abell himself (1958) observed that clusters were not
distributed uniformly on the sky, indicating a second-order
clustering.
Hauser \& Peebles (1973) calculated the two-point
angular correlation function $w_{cc}(\theta)$ of Abell clusters,
finding a strong correlation. Bahcall \& Soneira (1983) and
Klypin \& Kopylov (1983) calculated
the two-point spatial correlation function $\xi_{cc}(r)$ for
cluster samples with measured redshifts, finding
an amplitude $\sim 18$ times higher than that of galaxies.
Postman et al. (1986) recalculated $\xi_{cc}(r)$, considering the effect
of the Corona Borealis supercluster, and while finding a somewhat lower
amplitude confirmed the strong clustering of Abell (and Zwicky) clusters.

To have an all-sky catalog,
galaxy clusters with $\delta \le -17\degr$ have been independently selected
on deeper and more sensitive J plates (Abell, Corwin, Olowin, 1989).
The two-point angular correlation function $w_{cc}(\theta)$ of
ACO clusters has already been determined by Bahcall et al. (1988),
Couchman et al. (1988), Batuski et al. (1989; hereafter BBOB).
Also the three-point angular correlation function has been estimated
(T\'oth et al. 1989, Jing \& Zhang 1989).
Moreover, McGill \& Couchman have calculated the spatial correlation function
of $RC \ge 1$ ACO clusters by inverting $w_{cc}(\theta)$ --~assuming a model
for $\xi_{cc}(r)$~--, a technique subject to quite large uncertainties.

The above authors generally find comparable results for Abell
and ACO clusters for what concerns $w_{cc}(\theta)$ taken at its
{\em face value} --i.e., without correction for projection effects--.

Given the importance of the existence of structures
for the theories of galaxy formation, and given in particular the difficulty
of the CDM theory to account for the power at large scales
shown by the observed cluster distribution,
it was natural to ask if clusters are reliable
indicators of the large-scale structure.

Indeed there is presently a debate about the reliability of Abell and ACO
cluster catalogs. Projection effects might be important
(see Lucey 1983, Sutherland 1988, Dekel et al. 1989, Olivier et al. 1990,
Sutherland \& Efstathiou 1991): they can artificially
increase the amplitude of the 2-point correlation
function. However, Szalay et al. (1989) don't find such strong projection
effects, and X-ray selected clusters (Lahav et al. 1989),
and cD clusters (West \& van den Bergh 1991), which don't rely on
the same selection criterion of Abell clusters, show a strong
correlation.
Struble \& Rood (1991a) examined a sample of clusters with measured redshift
and found contamination only in 3\% of the 1682 Abell clusters
in the statistical sample. Moreover, Jing et al. (1992) find evidence that
real clustering and not contaminations can be the origin of the positive
redshift correlations  at large redshift  and small angular separations.

It is usually assumed that this problem will be clarified
by the availability of catalogs of clusters
derived from an automated search, like the recent APM survey of
galaxy clusters.
Nevertheless, it will be necessary
to assess how much these automated catalogs
are better than the classical, Abell and ACO catalogs,
and which kind of systems they are sampling --for example,
the APM survey contains only poor clusters, and this could
affect their correlations (Bahcall \&  West, 1992)--.

Very recently, not only $\xi_{cc}(r)$ has been subject to some
criticism, but also the result of Postman et al. (1992; hereafter
PHG92), who find $r_0 = 20.6$ \h ~for a sample of 208 clusters
with $\mid b_{II} \mid \ge 30\degr$, has
been questioned by Efstathiou et al. (1991; hereafter EDSM),
who find for the same data $r_0 = 17.4$ \h, and, after correcting for
projection effects, $r_0 = 13$ \h. This discrepancy between
PHG92 and EDSM for $r_0$ of clusters is important because,
quoting EDSM themselves, ``the differences in the respective estimates for
$\xi_{cc}(r)$ in the region where $\xi_{cc} \sim 1$ are similar to the
changes caused by the corrections for anisotropies".

In this paper we will
calculate and compare the ``observed" $\xi_{cc}(r)$ of Abell and ACO
clusters in two symmetric samples.
In this way we explore
different regions of the sky; this is an important test, because
the presence of
structures like Corona Borealis can affect the statistical analysis
(Postman et al. 1986).
We have also the opportunity of estimating the {\em spatial}
$\xi_{cc}(r)$ for a sample made only of ACO southern clusters.
This directly estimated ACO $\xi_{cc}(r)$ can be
compared with that of Abell clusters.

Moreover, in order to check if the spatial distributions of
clusters in the northern and southern galactic hemispheres are statistically
comparable we perform also cluster analysis
(Einasto et al. 1984, Tago, Einasto, Saar 1984),
and we search for southern superclusters, as an extension of the
PHG92 catalog.

In section 2, we describe our chosen samples and the reasons of this choice.
In section 3 we deal with their correlation functions; in section 4
we analyse their percolation properties
and we give a catalog of superclusters, with three different contrasts.
Our conclusions are in section 5.

In what follows, we assumed $H_0 = 100 h$ km sec$^{-1}$ Mpc$^{-1}$
and $q_0 = 0.1$, which are standard values used by other authors.

\section{The samples}

Up to recent times, a statistical analysis of clusters in the southern
galactic hemisphere has been impossible because of the $\delta$ limit
($\delta \ge -27\degr$) of Abell catalog.
The Abell, Corwin \& Olowin
(ACO, 1989) catalog of southern clusters has offered the opportunity to
extend our knowledge of the distribution of rich galaxy clusters.

In order to determine distances of clusters and
to study their internal dynamics,
a large quantity of cluster redshifts has been measured in the last years,
in both the northern and southern hemispheres.

We collected published redshifts for Abell clusters from Struble \& Rood
(1991b) and PHG92; for ACO clusters from
Muriel et al. (1990, 1991), Vettolani et al. (1989), Cappi et al. (1991),
and again PHG92 (who give redshifts of 15 ACO clusters).

We selected all clusters with $m_{10} \le 16.5$, $\mid b_{II} \mid \ge
40\degr$,
and richness class $RC \ge 0$.

The choice of $\mid b_{II} \mid \ge 40\degr$ allows us to have two complete,
equal-volume samples in the northern and southern galactic hemispheres.
Moreover, this symmetry minimizes any effect due to the galactic latitude
selection function.

The magnitude limit allows to avoid problems of contamination from
foreground / background galaxies, and to dispose both of a northern sample
where
all clusters have measured redshifts, and a southern sample where 90\% of
clusters have measured redshifts.
Of course, a magnitude limit corresponds to a flux limit; however, up to
z $\sim$ 0.08
cluster density is approximately constant, so we will limit our
samples at that redshift (see discussion in PHG92).

Our southern subsample (S40) includes 130 clusters with $b_{II} \le -40\degr$
and $m_{10} \le 16.5$; only 14 of them don't have any measured redshift.
For these 14 clusters we used the $log(cz) - m_{10}$
relation, as given by Scaramella et al. (1991, hereafter SZVC; see their
equation 1). In this
sample there are 60 Abell clusters and 70 ACO clusters.

In an analogous way
we extract from the PHG92 sample all clusters with $b_{II} \ge +40\degr$ and
$m_{10} \le 16.5$, with 103 clusters (N40). This represents a totally
symmetric northern sample.

We consider also a southern subsample (ACO) limited at $b_{II} \le -40\degr$,
$\delta \le -27\degr$, which includes 66 (all ACO) clusters.

Finally we reanalyse the PHG92 statistical sample (NST),
with 208 clusters ($\mid b_{II} \mid \ge 30\degr$, $\delta \ge -27\degr.5$,
$m_{10} \le 16.5$), for which
PHG92 and EDSM find discrepant results. Therefore we have decided to make
another, independent estimation of $\xi_{cc}(r)$ for this sample.

   \begin{table}
      \caption[ ]{Parameters of the samples
($z_{lim} \le 0.08$, $m_{10} \le 16.5$)}
       \label{TabPar}
	\begin{flushleft}
	\begin{tabular}{lllll}
            \hline
           &  N40 & S40 & ACO & NST \\
            \hline
$N_c$                  &  103     &  130     & ~66      & 208 \\
$z_{mean}$             & 0.0558   & 0.0566   & 0.0547   & 0.0566 \\
$n_c$ (h$^3$ Mpc$^{-3})$ & 1.1E-05  & 1.4E-05  & 1.6E-05  & 1.1E-05 \\
$\gamma$               & -1.90    & -1.71    & -1.71    & -2.02 \\
(+/--)                 &  0.42    &  0.33    &  0.53    &  0.37 \\
$r_0$                  & 19.4     & 19.7     & 20.1     & 19.2 \\
(+/--)                 & 5.1/4.2  & 3.3/2.8  & 6.6/5.1  & 3.5/3.0 \\
            \hline
            \end{tabular}
            \end{flushleft}
            \end{table}

In figures \ref{FigPoloN} and \ref{FigPoloS} we show the
projected distributions of N40 and S40 clusters, while
in fig. \ref{FigRed} we show their redshift distribution:
large structures are apparent.

We report in table 1 the main characteristics of each subsample.
Note the difference in cluster density of the northern and southern
subsamples, corresponding to a $3 \sigma$ poissonian deviation.
It has already been noticed from an analysis
of clusters with estimated distance $D \le 300$ \h ~that ACO clusters
have a higher density
(BBOB, SZVC). It is therefore interesting to verify if there are other
differences which can be revealed by statistical analysis.

\section{Correlation Analysis}

\subsection{Description of the method}

We calculated positions converting redshifts to distances in Mpc
using the Mattig formula for $q_0  > 0$ (Mattig, 1958; Weinberg 1972):

\begin{equation}
r = \frac{1}{q_0^2} \frac{c}{H_0} [1 - q_0 + q_0z + (q_0-1)
\sqrt{2q_0z+1}] / (1+z)
\end{equation}

where we assumed, as we noted previously, $H_0 = 100 h$ km $s^{-1}$
Mpc $^{-1}$ and $q_0 = 0.1$.

Clusters don't have high peculiar motions ($\le 1000$ km/s, Huchra et al.
1990).
Cluster redshifts can be considered reasonably accurate when at least
$\sim 4$ cluster galaxies have measured redshifts; however, even in the
case that only one galaxy has a measured redshift, this is usually the
brightest member, probably at the potential bottom of the cluster.

We used as estimator of $\xi_{cc}(r)$ the formula:

\begin{equation}
\xi_{cc}(r) = 2 \frac{N_{\rm ran}}{N_{\rm clu}} \frac{N_{cc}(r)}{N_{cr}(r)} - 1
\end{equation}

where $N_{cc}(r)$ is the number of cluster-cluster pairs at a distance $r$,
$N_{cr}(r)$ is the number of cluster - random point pairs
(this is the best way to avoid edge effects due to the limited size
of tested volume), $N_{\rm ran}$ is the total number of random points and
$N_{\rm clu}$ is the total number of objects.
Equation 2 is the standard estimator used to calculate the two-point
correlation function for both galaxies and clusters.
We assumed $N_{\rm ran} = N_{\rm clu}$, and
we generated 100 random catalogs each of them with the same
number of objects as that in the real sample, then averaging,
in order to have a good estimate of $N_{cr}(r)$.
Random objects were distributed uniformly taking into account
the non-euclidean geometry. Observed redshifts were assigned to
random points with a gaussian smoothing ($\sigma = 3000$ km/s)
to avoid possible effects of incompleteness in our
estimate of the correlation function.
The result is of course a lower amplitude than in the case we had assumed
no selection effect;
while if we did not have smoothed redshifts, some small-scale clustering would
have been reproduced in random catalogs, thus lowering the amplitude.
But the particular choice of $\sigma$ is not critical, as noted by PHG92,
if $3000 \le \sigma \le 6000$ km/s.

We adopted a logarithmic step ($\Delta$log(r) between $0.1$ and $0.2$
depending on samples). As a check, we recalculated
$\xi_{cc}(r)$ with a linear step (5 \h ~up to $r = 60$ \h,
then doubling the step to increase the signal to noise ratio).

Given the high cut in galactic latitude, the use of the selection function

\begin{equation}
P(b) = dex ~[C (1 - csc \mid b \mid)]
\end{equation}

where C~=~0.3 for Abell clusters and C~=~0.2 for ACO clusters,
has negligible effects. Anyway we estimated $\xi_{cc}(r)$ taking $P(b)$
into account.

We took into account also the $\delta$ selection function $P(\delta)$,
using the expression given by BBOB respectively for Abell clusters:

\begin{eqnarray}
P_{\rm Ab}(\delta) = & 1 & +24\degr \le \delta \le +90\degr \nonumber \\
              & 0.675 + 0.01125 \delta & -27\degr < \delta < +24\degr
\end{eqnarray}

and for ACO clusters:

\begin{equation}
P_{\rm ACO}(\delta) = dex ~[0.6 (cos \mid \delta \mid - 1)]
\quad -20\degr > \delta > -75\degr
\end{equation}

BBOB note that this effect may be partly due to real superclustering, partly
to a spur of galactic obscuration, or probably to higher air mass
at low declinations
(if this is the correct explanation, then SZVC show that it can be
described as a function of zenithal distance --see their equation 6--).

As in the case of $P(b)$, $P(\delta)$ does not change significantly
the results: the two above
selection functions represent second-order corrections.

We estimated errors through the bootstrap resampling method
(Ling, Frenk \& Barrow 1986);
these are the errors we show in graphs and used for
best-fitting.
The optimal way of determining errors can be matter of discussion, but
it is known that bootstrap errors are more realistic than poissonian errors.
We checked that best-fit parameters of $\xi_{cc}(r)$ do not change
significantly by using poissonian errors.

We fitted points with a least-square method, without imposing
any slope.
The choice of the range is important: values of $r_0$ and $\gamma$
depend on it. The correlation function of our samples drops
beyond 40 \h~ and below $\sim 5$ \h, deviating from the expected
power-law.
At small scales, below about 5 \h, we have
very few clusters at small separations --their small number
is attested also by the large error bars--.
Given that
clusters have linear Abell diametres of $3$ \h, below that scale
the correlation function is not meaningful.

It is also clear that we cannot significantly sample
pairs at scales comparable to the size of the sample,
i.e. typically beyond $\sim V^{1/3}/2$, where $V$ is the total volume.

We have chosen to fit data in the range $ 5 < r < 40$ \h,
where $\xi_{cc}(r)$ has a significant signal
--in that range it is always positive, being the first negative value
at 50 \h--.

In figures \ref{FigCorns}, \ref{FigCoraco} and \ref{FigCornst},
we report $\xi_{cc}(r)$ for our four samples, compared with
the power-law $\xi_{cc}(r) = (r/20)^{-2}$ (as EDSM did).
We assumed a logarithmic step $\Delta$log(r)=0.12 for all samples except
for the NST sample,
for which $\Delta$log(r)=0.1 (we will discuss below the consequences
of adopting a different step).
Uncertainties on $\gamma$ are derived from the
fit, while (1 $\sigma$) errors on $r_0$ are derived directly from
the same pseudo-data catalogs used for calculating bootstrap errors on
the correlation function.

\subsection{Discussion of results}

Our best-fit values for $\gamma$ and $r_0$ are shown in table \ref{TabPar}.
In our fixed range
slopes vary from -1.7 to -2.0 depending on samples.
They are all very similar (needless to say that the equal slopes of
the ACO and S40 correlations are a coincidence).

We find $r_0 \sim 19.4$ \h~ for
the N40 sample and $r_0 \sim 19.7$ \h~ for the S40 sample.
ACO clusters appear to have a correlation radius slightly
larger ($r_0 \sim 20.1 $ \h) and a flatter slope than
Abell clusters; anyway this small difference is less than $1 \sigma$,
and the presence of a very rich supercluster (see next section)
gives a significant contribution to the amplitude of
$\xi_{cc}(r)$ in this small sample.

To check the dependence of our results on the richness class,
we recalculated $\xi_{cc}(r)$ of clusters with $RC \ge 1$.
Of course, having only 49 northern in the N40 sample and 56 southern clusters
in the S40 sample, $\xi_{cc}(r)$ is more noisy; for that reason we have chosen
a logarithmic step $\Delta log (r) = 0.2$.
We obtain results which are fully consistent with those obtained
for the larger samples (fig.\ref{FigCorric}). Amplitude and slopes are
very similar to the total samples.
Therefore the southern sample S40 is a further confirmation of PHG92 result:
there is no clear richness effect in the distribution of nearby clusters,
at least between $RC = 0$ and $RC \ge 1$ clusters.

It is striking the similarity of the correlation
functions in the two opposite galactic hemispheres,
and of ACO and Abell clusters.
$r_0$ is much larger than that of galaxies, and consistent with other
published values of northern samples.
It is anyway smaller than the value of $30$ \h ~resulting from the indirect
estimate of McGill \& Couchman.

We detect in all subsamples
a positive signal only up to $45$ \h ~(for the estimate in linear step;
up to $\sim$ 40 ~for the logarithmic step): no superclustering is detected
beyond that scale.

We note that {\em all} the above samples show an anticorrelation between
50 and 100-120 \h, where $\xi_{cc}(r) \sim -0.1$. For all the 4 samples
--and also for the $R \ge 1$ N40 and S40 samples-- the first
negative point is at $r \sim 50$ \h.

In order to visualize this effect of anticorrelation at large scales
we plot $1 + \xi_{cc}(r)$ for our 4 samples (fig. \ref{FigCorxip}).
We do not report errors
to avoid confusion: it should already be clear to the reader that
all points are well within $1 \sigma$ error bars.

The strongest anticorrelation is shown by the N40 and NST samples
in the bin centered at $r \sim 60$ \h ~(respectively
$\xi_{cc}(r) \sim$ -0.3 and -0.14);
by the S40 and ACO samples in the next bin, centered at
$ r \sim 80$ \h ~(respectively $\xi_{cc}(r) \sim$ -0.18 and -0.22).
The effect is slightly more than 1 $\sigma$,
but the same behaviour is shown also by the N40 and ACO samples,
which are completely independent, and have different boundaries.
Calculating $\xi_{cc}(r)$ for the NST sample
with a cut in redshift of $z = 0.06$ and $z = 0.07$ we continue to find
the same effect with the strongest anticorrelation at the same value of r
($\sim 60$ \h).

This effect has been previously detected in the angular correlations:
Bahcall et al. (1988) and BBOB have shown that for two samples of
respectively Abell and ACO clusters
limited at an estimated distance $D \le 300$ \h, $w_{cc}(\theta)$
becomes negative ($\sim$ -0.18) at $\theta \sim 12\degr$, corresponding to
$\sim$ 50 \h. Their deeper samples do not show any anticorrelation.
They have suggested that this anticorrelation of nearby clusters might be due
to the presence of underdense regions between more clustered regions, and as
an example they indicate the region
at $l_{II} \sim 230\degr$ and $b_{II} \sim -50\degr$.
Alternatively, we note that it might be a spurious effect.
We do not know the ``universal"
mean density of clusters, therefore it is assumed $n = N_c / V_T$,
where $N_c$ is the total number of clusters and $V_T$ is
the volume of the sample.
The mean number of clusters at a distance r from a randomly chosen
cluster is

\begin{equation}
<N> = nV + n \int_0 ^r \xi_{cc}(r) dV
\end{equation}

where $n$ is the mean density (see Peebles, 1980).
When $r$ includes the whole sample,
the number of neighbors is $N_c-1$, while $nV = N_c$
with our choice of $n$. Then the above integral constraint implies
that $\xi_{cc}(r)$, being positive at small scales,
is forced to be negative at large scales.

Now we will investigate the reasons of the discrepancy between EDSM and PHG92.
In fig. \ref{FigCornst} we have shown $\xi_{cc}(r)$ for the PHG92 statistical
sample: we agree well with the results of PHG92.
Indeed, we do not find positive points between 50 and 100 \h,
where PHG92 find some positive signal, e.g. at $\sim 70$ \h.
This difference can be easily explained. Our bins are equal to those of PHG92
(0.1), but we have a different zero point. If we shift our zero point
of one half step we find
a positive point at $r \sim 70$ \h ~(see figure 3).
We verified that using $N_{rr}(r)$ --as PHG92-- instead of
$N_{cr}(r)$ there are no significant differences: $\xi_{cc}(r)$ is only
slightly higher especially at larger scales, where edge effects
become stronger (at the point at $\sim 70$ \h ~$\xi_{cc}(r)$ reaches
a level of $\sim 0.1$).
Our $r_0$ (19.2 \h) is smaller than that found by PHG92 (20.6 \h)
because they took into account all points up to 75 \h.
For example, fitting in a larger range, $1.5 < r_0 < 45$ \h,
we find $r_0 = 19.4$ \h ~and $\gamma = -1.89$, which are more
similar to the values found by PHG92.

Why then did EDSM find $r = 17.4$ \h ~and the last positive point at
$r \sim 30$ \h ~using a very similar estimator?

We are left with only two differences between PHG92 (and our) method and
that of EDSM: the redshift selection function and the logarithmic step.
The redshift selection function is however very similar, and it cannot
generate a significant difference.
On the contrary, we can show that here again the discrepancy is only apparent,
being due to a different logarithmic step.

We have recalculated $\xi_{cc}(r)$ trying to reproduce the positions and step
$\sim 0.2$, of EDSM (see their fig.4):
we have found $r_0 = 17.7$ \h, or an amplitude $10^{2.32 \pm 0.19}$
and a slope $\gamma = -1.86 \pm 0.17$, i.e. we obtained their actual
result with the same (poissonian) uncertainties.
The use of bootstrap errors does not change the fit but increases
substantially the uncertainties.
This is also visualized in fig.\ref{FigCornst} (asterisks connected
by a solid line). Notice
that the last positive point is at $\sim 31$ \h, because the following is
already at $r$ higher than 50 \h.
There is no intermediate point, therefore it
is not possible to appreciate the fact that correlation becomes negative
at $\sim 45$ \h, not at 30 \h. We believe that with that step there is a real
loss of information with respect to the smaller step.
Returning to fig. \ref{FigCornst}, it is possible
to appreciate that the best resolution is given by the 5 \h ~linear step.
In that case, a fit of $\xi_{cc}(r)$ gives $r_0 = 20.7$ \h.

As a useful exercise, we have recalculated $\xi_{cc}(r)$ of the S40 sample
with a logarithmic step $\Delta log (r) = 0.1$, instead of 0.12,
then we fitted points in the same range.
We find $r_0 = 19.1$ \h ~and $\gamma = -1.94$, to be compared to our previous
values $r_0 = 19.4$ \h ~and $\gamma = -1.71$.

These variations of the parameters induced by the chosen range and step
indicate clearly that values of $r_0$ and $\gamma$ as reported in table
\ref{TabPar} can only be rough estimates; what is really important is
the range of values, which is centered around $r_0 \sim 19.6$ and
$\gamma \sim -1.8$.

Moreover the above discussion makes clear that an accurate determination
of $\xi_{cc}(r)$ at scales $\ga$ 40 \h ~is not possible for the moment.
The problem is worse for deeper samples:
for example, Olivier et al. (1990) have found a discrepancy at large scales
between angular correlations of Abell and ACO clusters of distance classes
5 and 6. Their corrected $w_{cc}(\theta)$ indicates a null correlation for
ACO clusters at large scales.

\subsection{Cluster pair elongations}

High peculiar motions and/or projection effects can cause an
elongation of cluster pairs in the redshift direction.
To check for the presence of elongations, we can study the
distribution of values of the
angle $\beta$ between the line which connects a pair and
the line-of-sight direction --or
the angle $\alpha = 90\degr - \beta$ formed with the plane of the sky--
as a function of the spatial separation of clusters. This test, proposed by
Sargent \& Turner (1977), has been used by PHG92, who didn't detect
significant elongations in their statistical sample.
We measure $\beta$ not at the midpoint between the pair,
as in Sargent \& Turner, but at the midpoint of the projected separation,
therefore we have:

\begin{equation}
tan(\beta) = tan \left( \frac{\theta}{2} \right) \frac{D_1 + D_2}{D_1 - D_2}
\end{equation}

where $\theta$ is the angular separation of the two clusters,
and $D_1$ and $D_2$ are their respective distances ($D_1 > D_2$);
because of geometrical constraints, $\beta$ is always $ > \theta/2$.
It is important to eliminate all pairs whose distance from
the limits of the sample is less than their separation, otherwise
the distribution would be biased in the redshift direction.
In fig.\ref{Elongation}a-d we report the histograms of the distribution
of cos($\beta$) for the N40 and S40 samples corresponding to different
spatial separations.

All the distributions are similar
(the distribution of cos($\beta$) for clusters within 15 \h ~
is comparable to that for clusters with larger separations) thus
indicating that there are no significant elongations in the samples.
Moreover, the mean value of $\alpha$ is very similar at all
separations in both samples, $\sim 35\degr$, to be compared to
the value $32.7\degr$ expected for an isotropic distribution.

\section{Cluster analysis}

\subsection{Percolation analysis}

In order to verify the similarity of Abell and ACO cluster
spatial distributions, we applied cluster
analysis (Einasto et al., 1984; Tago, Einasto, Saar, 1984).

We searched for all clusters with separation less than
a fraction $s$ (the percolation parameter)
of the mean inter-cluster distance $r_m = n^{-1/3}$.
A set of clusters where each cluster is at a separation less than $s$ from
another cluster constitutes a supercluster.
Superclusters with at least three members give an additional,
higher-order information relatively to $\xi_{cc}(r)$.

Dekel \& West (1985) showed that percolation
has some problems: they can be reduced by using volume-limited
samples confined to the same volume, and this is substantially
the case for our N40 and S40 samples, to which we will limit our
analysis below.

Figure \ref{FigPer} shows that percolation of
the two samples is similar. We plot $l_p = L_{max} / L_s$, where $L_{max}$
is a measure of the largest supercluster (taken as the maximum distance
between two clusters in the same supercluster) and $L_s$ is the size of
the sample (defined
as $V^{1/3}$; this is not a critical definition, because the two
samples have identical volumes), as a function of $s$.
We report again bootstrap errors (for sake of clarity, only errors
for S40 sample are displayed; errors for N40 are comparable), computed from 100
pseudo-data catalogs.

In figures \ref{FigAlpha}, \ref{FigBet} and
\ref{FigGam} we visualize the parameters $\alpha$,
$\beta$ and $\gamma$: they represent the fraction of clusters in small,
intermediate and large superclusters. They are found by dividing the maximum
multiplicity $m_{max} = log_2(N_c)$, where $N_c$ is the total number of
clusters in the sample, into three equal parts. $\alpha$ includes isolated
clusters, so that $\alpha + \beta + \gamma = 1$.

At first sight, no main difference is apparent.
The two sample show an identical behaviour up to $s = 0.6$
(corresponding to a scale of $\sim 25$ \h ~for the southern
sample and $\sim 31$ \h ~for the northern sample).
Beyond $s = 0.6$ there is indeed some difference.
The S40 sample percolates before the N40 sample (at $s=0.95$
vs. $s=1.09$);
moreover, it has a lower fraction of clusters in small superclusters
($\alpha$ curve) and a higher fraction in rich
superclusters ($\beta$ and $\gamma$ curves).
However, the difference is at the $1 \sigma$ level.
The fact that it appears beyond $s=0.6$
means that we are examining large structures at a very low
contrast.
It is probably due to the presence of a rich
supercluster in the south galactic hemisphere, which we will describe
in the next section.

\subsection{Superclusters}

We searched for superclusters
selected at different contrast, mainly as
a ``complement" of PHG92, in its turn an extension of
Bahcall \& Soneira catalog (1984; see also Batuski \& Burns, 1985b).
We do not have the limit  $\delta = -27\degr 30'$,
but a higher limit in $b_{II}$.
We consider separately the two galactic hemispheres,
using the N40 and S40 samples.
For the lowest contrast,
we have fixed $s = 0.5$; given the different densities,
it corresponds to a length $r_s = 20.6$ \h~\ for SA and $r_s = 22.3$ \h \
for NA and, from the formula

\begin{equation}
\frac{\delta n}{n} = ({4 \over 3} \pi r_s^3 n)^{-1} - 1
\end{equation}

to a contrast of $\sim 0.9$ or
a space density enhancement $f = 1 + \delta n/n \sim 1.9$.
We have considered also the enhancements $f=5$ and $f=10$
(corresponding to $s = 0.363$ and
$s = 0.288$) to allow a direct comparison with PHG92 results.
In tables ~\ref{TabSupN} and ~\ref{TabSupS} we give a catalog
of these superclusters for each of the chosen enhancements.
Of course, many superclusters are common to those found by PHG92
(see their table 5), mainly the northern ones.
We missed some superclusters because of the higher $b_{II}$ limit.
We have one more supercluster, N6, simply because our enhancement is
1.9 and not 2.0 --with the higher enhancement
the cluster A1709 would not be connected to the system--.
The most relevant features in this north galactic cap are N9,
which corresponds to the well known Hercules cluster
-- A2197/A2199 region and the Corona Borealis
supercluster (N10) (see Giovanelli \& Haynes, 1991).
In the south galactic cap we find 4 more superclusters, S10, S11, S12, S13.
S11 is particularly interesting: it is made of 15 clusters
(7 with $RC = 0$) at the lowest
contrast, and it is broken into two subsystems at the higher
contrasts. Its characteristic size, defined as the largest separation
between two members, is 65 \h ~($f = 1.9$), and
its mean redshift is $\sim 0.06$. Moreover, it could extend beyond our
limits in $b_{II}$ and $z$. We cannot define its precise extension
also because 6 of its 15 members happen to have an estimated distance
(4 are $RC = 0$ clusters). Anyway this structure is surely real,
as it was demonstrated by Lucey et al. (1983) mainly on the
basis of the galaxy distribution, and
it is prominent in the cone diagram that
we show in fig. \ref{FigCono}, where it appears as a filament
perpendicular to the line of sight, in front of a concentration at
higher redshift; these two parts
are separated at high contrast (respectively S11b and S11a in table
\ref{TabSupS}).
Lucey et al. called it the Horologium-Reticulum supercluster; it is in a
region characterized by a higher density of galaxies with
$v \sim 18000$ km/s (see also Giovanelli \& Haynes, 1991).
This large structure traces partially the edge of a large void.
S9 is the central part of the Pisces-Cetus supercluster candidate
of Batuski \& Burns (1985b).
Here again our catalog does not include some systems
because of their low galactic latitude.

It is again remarkable the similarity between the two hemispheres.
At the lowest contrast
there are 19 superclusters in N40 and 20 in S40, taking into account
also binary systems. Largest superclusters are 63 \h~\ in N40 and
65 \h~\ in S40.
The numbers of superclusters with more members are
similar. The southern and northern
richest superclusters have respectively 15 and 10 members.
A total of 69 clusters are in systems of 2 or more clusters in N40;
84 clusters are in corresponding systems in S40.
This means that 67\% of northern clusters and 65\% of southern clusters
are in systems with 2 or more members; these percentages become
respectively 51\% and 54\% if we count only superclusters with at least
3 members.

If we choose a higher density enhancement, evidently we find
less superclusters with less members. However,
even for an enhancement of 10, we continue to find
17 and 16 superclusters with at least 2 members respectively
in the northern and southern hemispheres;
the richest ones have 4 clusters. These are
concentrations of clusters, which at lower contrast are
connected to other clusters. At the higher contrasts there is a
larger number of rich
superclusters in the S40 sample. This is partly due to the presence of
two systems, S1 (9 members at $f = 1.9$) and the already discussed S11,
which are broken into two smaller systems at higher contrasts.

The characteristic sizes of superclusters (including binary systems, not
reported in the tables) are visualized on the histograms in
fig.\ref{FigSizN} and
\ref{FigSizS}, fixing a space density enhancement $f = 1.9$.
The dashed line represents the expected distribution for a random
sample with the same number of objects and the same volume as in the real
sample
(we made 50 random catalogs and averaged the results).
A comparison with the corresponding random samples is needed because
of the difference in density between N40 and S40.
It appears that the N40 and S40 distributions are similar.
The most common size of superclusters is in the range
20 - 30 \h, while for random catalogs we would expect it in
the range 10 - 20 \h; on the contrary, in this second range
real samples have less superclusters than the poissonian expectation.
It is well apparent the excess of large structures,
up to $\sim 60$ \h, and of rich systems
(figures \ref{FigRicN} and \ref{FigRicS}),
which are not present in random samples.

We come to the conclusion that the two samples have very similar
distributions from a statistical point of view:
the main difference is their density.
Moreover the southern polar cap presents a particularly rich
supercluster.

As a final observation, we note that our samples do not include
the so-called Shapley concentration
(Raychaudhury 1989, Scaramella et al. 1989) made (with one exception)
by ACO clusters at northern galactic latitude but below our limit
$b_{II} \ge +40\degr$.

\begin{table}
\caption[ ]{Northern superclusters $b_{II} \ge +40\degr$, $z \le 0.08$}
\label{TabSupN}
\begin{flushleft}
\begin{tabular}{llll}
\hline
   ID & No. & Members (enhancement $f$ = 1.9) & PHG ID \\
\hline
    N1  &  4 &  999, 1016, 1139, 1142 & 2 \\
    N2  &  3 & 1149, 1171, 1238 & 3 \\
    N3  &  5 & 1177, 1185, 1228, 1257, 1267 & 4  \\
    N4  &  3 & 1216, 1308, 1334 & 5 \\
    N5  &  8 & 1270, 1291, 1318, 1377, 1383, & 6 \\
        &    & 1436, 1452, 1507 &  \\
    N6  &  3 & 1631, 1644, 1709 & \\
    N7  &  4 & 1775, 1800, 1831, 1873 & 7 \\
    N8  &  3 & 1781, 1795, 1825 & 8 \\
    N9  & 10 & 2052, 2063, 2107, 2147, 2148, & 9  \\
        &    & 2151, 2152, 2162, 2197, 2199  &    \\
   N10  &  7 & 2061, 2065, 2067, 2079, 2089, & 10 \\
        &    & 2092, 2124                    &    \\
   N11  &  3 & 2168, 2169, 2184, & 11 \\
\hline
 ID & No. & Members (enhancement $f$ = 5) & PHG ID \\
\hline
    N3    &  5   &  1177, 1185, 1228, 1257, 1267  & 4 \\
    N5    &  6   &  1291, 1318, 1377, 1383, 1436, & 6 \\
          &      &  1452 & \\
    N7    &  4   &  1775, 1800, 1831, 1873 &  7 \\
    N9    &  5   &  2107, 2147, 2148, 2151, 2152 & 9 \\
   N10    &  4   &  2061, 2065, 2067, 2089 & 10 \\
\hline
 ID & No. & Members (enhancement $f$ = 10) & PHG ID \\
\hline
    N3    &  3 &  1177, 1185, 1267 & 4 \\
    N5    &  4 &  1291, 1318, 1377, 1383 & 6 \\
    N7    &  4 &  1775, 1800, 1831, 1873 & 7 \\
    N9    &  3 &  2147, 2151, 2152 & 9 \\
\hline
\end{tabular}
\end{flushleft}
\end{table}

\begin{table}
\caption[ ]{Southern superclusters ($b_{II} \le -40\degr$, $z \le 0.08$)}
\label{TabSupS}
\begin{flushleft}
\begin{tabular}{llll}
\hline
 ID & No. & Members (enhancement $f$ = 1.9) & PHG ID  \\
\hline
    S1  &  9 &   14,   27,   74,   86,  114  & 13 \\
        &    &  133, 2716, 2800, 2824        &    \\
    S2  &  3 &   85,  117,  151              & 14 \\
    S3  &  3 &  102,  116,  134              & 15 \\
    S4  &  6 &  119,  147,  160,  168, 193,  195 & 16  \\
    S5  &  3 &  154,  158,  171  & 17 \\
    S6  &  4 &  419,  428, 3094, 3095        & 18 \\
    S7  &  3 & 2366, 2399, 2415 & 20 \\
    S8  &  3 & 2459, 2462, 2492 & 21 \\
    S9  &  4 & 2589, 2592, 2593, 2657, & 22 \\
   S10  &  6 & 2731, 2806, 2860, 2870, 2877, 2911 &  \\
   S11  & 15 & 3089, 3093, 3100, 3104, 3108, & \\
        &    & 3111, 3112, 3122, 3123, 3125, & \\
        &    & 3128, 3133, 3135, 3158, 3164  & \\
   S12  &  4 & 3144, 3193, 3202, 3225        & \\
   S13  &  7 & 3771, 3785, 3806, 3822, 3825  & \\
        &    & 3826, 3886  & \\
\hline
 ID & No. & Members (enhancement $f$ = 5) & PHG ID \\
\hline
    S1a   &  3   &    74,   86, 2800   &  13 \\
    S1b   &  3   &   114,  133, 2824  &  13 \\
    S4    &  3   &   119,  147,  168  &  16 \\
    S6    &  3   &   419, 3094, 3095 &  18 \\
    S9    &  3   &  2589, 2592, 2593  &  22 \\
   S10    &  5   &  2731, 2806, 2860, 2870, 2877 &  \\
   S11a   &  7   &  3093, 3100, 3104, 3108, 3111, & \\
          &      &  3112, 3133 &  \\
   S11b   &  4   &  3125, 3128, 3158, 3164 &  \\
   S12    &  4   &  3144, 3193, 3202, 3225 &  \\
   S13    &  4   &  3806, 3822, 3825, 3826 &  \\
\hline
 ID & No. & Members (enhancement $f$ = 10) & PHG ID \\
\hline
   S1a &   3 &   74,   86, 2800 & 13 \\
   S1b &   3 &  114,  133, 2824 & 13 \\
   S4 &    3 &  119,  147,  168 & 16 \\
  S10 &    3 & 2860, 2870, 2877 &    \\
  S11a &   3 & 3104, 3111, 3112 &    \\
  S11b &   3 & 3125, 3128, 3158 &    \\
  S13  &   4 & 3806, 3822, 3825, 3826 & \\
\hline
\end{tabular}
\end{flushleft}
\end{table}

\section{Conclusions}

We compared the clustering
properties of two complete, high galactic latitude cluster samples symmetric
to the galactic plane and we made a direct estimate of the two-point spatial
correlation function of ACO clusters in the southern galactic
hemisphere. From the analysis of these samples
we can draw the following conclusions.

\begin{itemize}
\item Correlation functions are all compatible with the same power-law
$\xi_{cc}(r) = ({r\over{r_0}})^{-\gamma}$.
In the fixed range $5 < r < 40$ \h ~we find consistent values
of $r_0$, $19.3 \le r_0 \le 20.1$ for all subsamples.
The largest value corresponds to the ACO clusters, but the difference
is not statistically significant. From the northern and southern
symmetric sample we derive a mean value of 19.6 \h ~for $r_0$.

\item In the same range of $r$
slopes have values from -1.7 to -2.0.
The northern samples
have a steeper slope, but here again, the difference is not statistically
significant (1 $\sigma$).
The mean slope of the northern and southern clusters is $\gamma \sim$ -1.8.

\item For the statistical sample NST we find $r_0 = 19.2$ and
$\gamma = -2.0$. These values are a little different from those
given by PHG92 because of the different fit range.
We have shown that EDSM have found different results mainly because of their
larger step. We have shown that it is important to use a sufficiently
small step to determine where $\xi_{cc}(r)$ becomes negative.

\item We find for $r > 40$ \h~a quite steep cutoff
for both northern and southern clusters; the first negative
value is at $ r \sim 50$ \h~.
We find a small (significative at the $1 \sigma$ level) but systematic
(it is present in all samples) anticorrelation between
$50$ and $100 - 120$ \h, with minima between $-0.14$ and $-0.3$
depending on samples;
this effect had been previously detected by Bahcall et al. (1988)
and BBOB analysing the angular correlation functions.

\item Percolation properties are similar in the northern and
southern hemispheres, but the S40 sample has richer structures
for values of the percolation parameter $s >0.6$.

\item We have given catalogs of superclusters respectively for the
northern and southern galactic hemispheres, mainly
as a supplement to the PHG92 catalog for the southern galactic
hemisphere. The main feature at high latitude in the southern galactic
hemisphere is a large supercluster (15 members) in the Horologium region,
at a mean redshift $z \sim 0.06$, with a linear size $D \sim 65$ \h,
near to a large void (see Lucey et al., 1983).

\end{itemize}

The above results do not not depend on the galactic latitude or
delta selection functions $P(b)$ and $P(\delta)$;
exclusion of richness 0 clusters does not change our results.
Therefore our general conclusion is that nearby Abell and ACO
clusters both in the north and south galactic hemispheres
have a much higher correlation function than galaxies;
their $\xi_{cc}(r)$ becomes negative beyond $\sim 45$ \h.
Correlations of clusters in the northern and southern galactic
hemispheres and their clustering properties don't show any difference
which can be claimed statistically significant.
Of course, this implies that ACO southern clusters have the same
distribution as Abell clusters: if there are important projection effects,
then both of them are affected in the same way. This is confirmed
by the spatial correlation function of the small sample of ACO clusters.

Finally we remark that other valuable information on cluster distribution
can be obtained from the scaling properties of the Void Probability Function
(Jing, 1990; Cappi, Maurogordato, Lachi\`eze-Rey 1991) and from counts in cells
(Cappi \& Maurogordato, 1991). The application of these statistics to
the available cluster samples will be the object of a following paper.

\acknowledgements
We thank M.Lachi\`eze-Rey for very useful discussions, M. Postman and
M.F. Struble for providing us their data before publication.

   \begin{figure}
      \caption[]{N40 sample (outer contour: $b_{II} = +40\degr$)
       \label{FigPoloN}}
    \end{figure}
   \begin{figure}
      \caption[]{S40 sample (outer contour: $b_{II} = -40\degr$)
       \label{FigPoloS}}
    \end{figure}

   \begin{figure}
      \caption[]{Redshift histogram (solid line: SA clusters;
dashed line: NA clusters) \label{FigRed}}
    \end{figure}
   \begin{figure}
      \caption[ ]{The spatial correlation function $\xi_{cc}(r)$ for the
N40 sample (asterisks); S40 sample (circles);
dashed line: $\xi_{cc}(r)=(r/20)^{-2}$ \label{FigCorns}}
    \end{figure}
   \begin{figure}
      \caption[ ]{$\xi_{cc}(r)$ for the ACO sample; dashed line
as in fig. \ref{FigCorns} \label{FigCoraco}}
    \end{figure}
   \begin{figure}
      \caption[ ]{$\xi_{cc}(r)$ for the NST sample;
asterisks: $\Delta$log(r)=0.2; circles: $\Delta$r=5 \h;
crosses with error bars: $\Delta$log(r)=0.1;
dashed line as in fig.\ref{FigCorns} \label{FigCornst}}
    \end{figure}
   \begin{figure}
      \caption[ ]{$\xi_{cc}(r)$ for the N40 (asterisks) and S40 (circles)
rich clusters ($R \ge 1$); dashed line as in fig.\ref{FigCorns}
       \label{FigCorric}}
    \end{figure}
   \begin{figure}
      \caption[ ]{$1 + \xi_{cc}(r)$: asterisks: N40; circles: S40;
crosses: ACO; squares: NST; dashed line as in fig.\ref{FigCorns}
       \label{FigCorxip}}
    \end{figure}
%

   \begin{figure*}
      \caption[ ]{Histograms of cos($\beta$) distribution.

{\bf a} N40 sample.
Solid line: pairs with separation $ D_{sep} < 15 $\h; dashed line:
pairs with separation $15 \le D_{sep} < 30 $ \h;~
{\bf b} N40 sample.
Solid line: pairs with separation $ 30 \le D_{sep} < 45 $\h; dashed line:
pairs with separation $45 \le D_{sep} < 60 $ \h;~
{\bf c} S40 sample.
Solid line: pairs with separation $ D_{sep} < 15 $\h; dashed line:
pairs with separation $15 \le D_{sep} < 30 $ \h;~
{\bf d} S40 sample.
Solid line: pairs with separation $ 30 \le D_{sep} < 45 $\h; dashed line:
pairs with separation $45 \le D_{sep} < 60 $ \h~ \label{Elongation}}
    \end{figure*}

   \begin{figure}
      \caption[ ]{$l_p$ as a function of $s$. Solid line: S40; dashed line:
N40 \label{FigPer}}
    \end{figure}
   \begin{figure}
      \caption[ ]{$\alpha$ as a function of $s$. Solid line: S40; dashed line:
N40}
         \label{FigAlpha}
    \end{figure}
   \begin{figure}
      \caption[ ]{$\beta$ as a function of $s$. Solid line: S40; dashed line:
N40}
         \label{FigBet}
    \end{figure}
   \begin{figure}
      \caption[ ]{$\gamma$ as a function of $s$. Solid line: S40; dashed line:
N40}
         \label{FigGam}
    \end{figure}

   \begin{figure}
      \caption[ ]{Slice of the S40 sample: $210\degr \le l_{II} \le 300\degr$,
$-75\degr \ge b_{II} -40\degr$, $z \le 0.08$}

         \label{FigCono}
    \end{figure}
   \begin{figure}
      \caption[ ]{Size distribution of superclusters (solid line:
N40 sample; dashed line: random sample)}

         \label{FigSizN}
    \end{figure}
   \begin{figure}
      \caption[ ]{Size distribution of superclusters (solid line:
S40 sample; dashed line: random sample)}

         \label{FigSizS}
    \end{figure}
   \begin{figure}
      \caption[ ]{Richness distribution of superclusters
(solid line: N40 sample; dashed line: random sample)}

         \label{FigRicN}
    \end{figure}
   \begin{figure}
      \caption[ ]{Richness distribution of superclusters
(solid line: S40 sample; dashed line: random sample)}
         \label{FigRicS}
    \end{figure}

\end{document}